\documentstyle[amscd,amssymb,verbatim,12pt]{amsart}
\def\dspace{\baselineskip=0.3 in}
\begin{document}
\dspace
\title[VARYING GRAVITATIONAL CONSTANT, ...  ]{VARYING GRAVITATIONAL CONSTANT AS WELL AS COSMOLOGY FROM THE
  EARLY INFLATION TO LATE ACCELERATION AND FUTURE UNIVERSE}

\author{\bf S.K.Srivastava}
{ }
\maketitle
\centerline{ Department of Mathematics,}
 \centerline{ North Eastern Hill University,}
 \centerline{  Shillong-793022, India}
\centerline{ srivastava@@nehu.ac.in; sushilsrivastava52@@gmail.com}

\vspace{1cm}

\centerline{\bf Abstract}

Here, cosmology is obtained from the variable gravitational constant $ G \propto \phi^{-2}$ with $ \phi(x) $ being a scalar and its fluctuations around the ground state. The gravitational action contains Einstein-Hilbert like term with variable $ G $, kinetic energy and self-interaction potential for $ \phi(x) $. Two phase transitions take place in this model. The first one takes place at the GUT (grand unified theory) scale $ \sim 2.45 \times 10^{14}{\rm GeV} $, when the early universe exits the inflationay phase and the second one at the electro-weak scale. Spontaneous symmetry breaking takes place around this scale As a consequence, variable $ G $ acquires constant value $G_N$ (the Newtonian gravitational constant).The standard model of cosmology is obtained in the post-second phase transition era. Interestingly, the dark matter and quintessence dark energy are created from the gravitational sector as a combined effect of the linear term of scalar curvature and $ \phi(x) $ without using non-linear terms of curvature or any other matter fields. Radiation-dominance is obtained in  both the pre-second phase transition era and the post-second phase transition era. In the pre-second phase transition era after inflation, radiation dominated phase of the universe decelerates in the non-standard manner as $\sim t^{1/3} $, but expands in the standard way as $\sim t^{1/2} $ in the post-second phase transition era. It is interesting to find that fluctuations in $ \phi(x) $ cause the phantom dark energy after the second phase transition, which grows later in the expanding universe. Dominance of dark matter over radiation is obtained at red-shift 1091.89. Matter-dominated phase decelerates as $\sim t^{2/3} $. Phantom dark energy, created during radiation-dominance of the post-second phase transition era, overpowers cosmic dynamics at red-shift 0.46. The late universe accelerates driven by the phantom dark energy, when time $t > 0.13 t_0 $ (where $ t_0 $ is the present age of the universe). It is found that phantom dark energy evaporates ultimately contributing to the vacumm energy with density almost twice of the present dark energy density. It is also found that the phantom era is free from big-rip singularity. 



\vspace{1cm}

\centerline {\underline{\bf 1. Introduction}}
 \smallskip

Observation of the late cosmic acceleration \cite{sp, ag}, made during last few
years, 
challenged the standard model of cosmology predicting deceleration in the
present universe. This revolutionary result (late cosmic acceleration)
prompted cosmologists to think dominance of some exotic
fluid in the very late universe. Theoretically, the basic characteristic of such a fluid is obtained from Freidmann equations giving the cosmic dynamics. These equations suggest negative pressure $p < - (1/3)\rho $ with $\rho $ being the energy density) for the dominating fluid to get the late cosmic acceleration. This requirement implies violation of the {\em strong
  energy condition} (SEC). Being unknown and having gravitational effect, this exotic fluid is popularly known as {\em dark energy }(DE). Later on, the case $p < - \rho $
was found better fit to observations. In this case, the {\em weak energy condition}(WEC) is also violated. 
This type of dark energy fluid was dubbed {\em phantom} \cite{cald}. Apart from these models,
many other field-theoretical and hydro-dynamical models of DE were proposed in
the recent past to explain challenging observation of late acceleration. A
review of these investigations is available in \cite{ejc}. In the race to
search  models to explain the phenomenon of late cosmic acceleration as well as to satisfy solar
system constraints, curvature was proposed as gravitational alternative of
DE. This work is reviewed in \cite{snj}. These models, specially model with non-linear terms $ R^n $ and $ R^{-m} $ \cite{sn03}received  criticism
in \cite{lds} on the ground 
that these models  did not produce matter in
the late universe needed for formation of large scale structure in the
universe. In another review, Nojiri and Odintsov  discussed dark matter in
the late universe responding to criticisms against their work \cite{sn08}. In
refs. \cite{snj, lds, sn08}, non-linear curvature terms are treated as DE term 
{\em a priori} and its consequences are discussed.  But, instead of assuming
curvature as DE term  it is more appropriate
to know whether curvature can produce DE density terms in the Friedmann
equation without assuming  non-linear curvature terms  as DE term 
{\em a priori}. This aspect is addressed to in \cite{sks, sks06, sks07, sks08a,  sks08b, sks08c} using an approach {\em different} from the approach of
refs.\cite{snj, lds, sn08}. In this approach, matter is created from gravity itself, which resolves the problem raised in \cite{lds} against modified gravity .

In spite of so many efforts to investigate an appropriate model of DE, it is not unreasonable to think for
some other alternative, which is planned to probe in what follows. In this context, it is useful to consider our past experiences. We know that Kaluza-Klein (K-K) modes are not excited  below a TeV scale. According to Ref. \cite{nah}, string/M-theory excitations are also expected at a TeV scale. Moreover, coupling of K-K modes to matter or gravitational excitations is set by the weak scale. Electromagnetism and weak interaction are unified at the scale 246 GeV, which is observed. Although at CERN's LHC 
it is possible to reach the energy scale $10^4 {\rm GeV}$, we have a renormalizable theory at 246 GeV in the observed sector (GUT scale is not observed) of mass scale.

So far, gravity is observed much below this scale. One possible reason for non-observation of gravity at high energy scale is the effect of different non-gravitaional forces at high energy, which are almost negligible at low energy.

These arguments, which are based on experimental and theoretical results, prompt us to think deviation in the gravitational constant $ G = G_N $ . One possible deviation is the space-time dependence of $G$. On the basis of above arguments, this type of deviation is proposed here above the electroweak scale 246 GeV. Another motivation for variable $G$ is given by the  
possibility of physical constants to have different values in the remote past of the cosmic history. An evidence for this possibility is given by the apparent observation that fine structure constant had different value in the distant past \cite{jkw}.

This type of proposal is not new in the history of gravity. Varying $ G $ had been considered by many authors in the past \cite{pj, bd, pgb, kn,rvw, jdb}. First time, it was proposed in Ref.\cite{pj} by Jordon in his scalar-tensor theory taking $ G $ proportional to an arbitrary power of $\phi^{-1} $ with $\phi(x,t)$ being a scalar. This theory was developed later by Brans and Dicke in Ref.\cite{bd} using $ G \propto \phi^{-1} $. In these theories, the kinetic energy term for $\phi(x,t)$ is non-canonical having the form $ \omega \partial^{\mu}\phi \partial_{\mu}\phi/\phi$ with $ \omega $ being a coupling constant. Later on, in Refs.\cite{pgb, kn,rvw,jdb}, these theories were generalized considering $ \omega $ dependent on $ \phi $.

In the old gravitational theories with variable $ G $ (OGTVG) proposed in the past \cite{pj, bd, pgb, kn,rvw,jdb}, $G $ is {\em not equal} to $G_N$ even in the present universe. But experiments support $G = G_N$ in the present universe as results of Einstein's theory of gravity, a theory with $G = G_N $, support observations of advance of perihelia and bending of light rays around the Sun satisfactorily ,whereas the Brans-Dicke theory which is the most popular theory having variable $G $ in the present universe too has problem in explaining advance of perihelia and bending of light rays. Astrophysical observations, made at the dawn of this century \cite{sp, ag} concluding late cosmic acceleration as well as content of the present universe in the form of $73\%$ dark energy and $23\%$ dark energy, are based on $ G = G_N $ in the late universe.

Moreover, on taking $G \sim \phi^{-1}$ the kinetic energy term for $\phi $ is non-canonical. So, a theory is needed to explain variable $G$ in the past and  $G = G_N$ in the present universe.

It is possible to have a theory with variable $G$ at high energy scales but $G = G_N $ in the late universe and canonical kinetic energy term for $\phi$ , if $ G \sim \phi^{-1}$ is replaced by $ G \sim \phi^{-2}$ with potential 
$V(\phi) = \lambda(\phi)|\phi^2 - \beta^{-1}M_P^2|^2/4,$ 
 where dynamical coupling $\lambda \propto \phi^{-2}$ is dimensionless,  $M_P$ is the Planck mass and $\beta $ is a dimensionless coupling constant. This is possible without causing any harm to the theory. 
 
 On the basis of above experimental probes getting a renormalizable theory upto the electro-weak scale $246$ Gev in the particle sector and probe of gravity upto $0.1$mm., it is proposed here that phase transition takes place at $\phi = \beta^{-1/2}|M_P |$ giving $ G = \beta \phi^{-2} = M_P^{-2} = G_N$ at the electro-weak scale. But prior to the phase transition, $ G = \beta \phi^{-2}$ ie. above the energy scale $246$ Gev, $G$ is variable.

 The present model is different from OGTVG proposed in the past \cite{pj, bd, pgb, kn,rvw,jdb}in the following four different ways. 
 
 \noindent (i) $G \propto
 \phi^{-2}$ in the present model, whereas $G\propto
 \phi^{-1}$ in OGTVG except Jordon's theory with an arbitrary power of $\phi^{-1}$. 
 
 \noindent (ii)In the present model, $\phi$ has canonical kinetic energy term and quartic self-interaction potential
 with a dynamical coupling. This set-up is in contrast to OGTVG having non-canonical kinetic energy term and no potential for $\phi$.

 \noindent  (iii) In the present model,$G $ is a variable above the electro-weak scale and this is equal to $G_N$ in the present universe. This set-up is in contrast to OGTVG
 , where $G$ is always a variable. 
 
 \noindent  (iv) Mass dimension for $ \phi $ is {\em one}in the present model, whereas mass dimension of $ \phi $ is {\em two} in OGTVG (in natural units given below).

\noindent {\bf Set up of the model and consequences}.  Here, the
 gravitational action contains Einstein-Hilbert like term with $G = \beta\phi^{-2}$ (with $ \beta $ being a dimensionless coupling constant),
 kinetic energy for $\phi$ (required for $\phi$ to be a dynamical variable) and self-interaction potential $V(\phi)$. $\phi$ is extremely small at the Planck
 scale and increases with the expansion of the universe. It is obtained that, in the flat, homogeneous and isotropic model of the universe, trace of gravitational field equations derived from the considered action yields the Friedmann equation with dark matter density and energy density of an unknown fluid, which mimics quintessence dark energy and derives early inflation. These energy density terms emerge from the gravitational sector ( mentioned above) without taking any other source except $ \phi $ and scalar curvature $ R $. 

Using observational consequences, it is found that universe exits from the early inflation at $ \phi = \phi_c = 3.45\times 10^{-13} M_P$ expanding around $ 10^{27} $ times. 

Two phase transitions take place in this model. The 
 The first one takes place at $ a = a_c $ (when $ \phi = \phi_c $) and enormous amount energy is released as latent heat. The emitted radiation (latent heat) heats up the universe upto the temperature $ T_c = 2.45 \times 10^{14}{\rm GeV}$. 
Due to creation of this radiation , the derived Friedmann equation gets modified. It is found that the emitted radiation dominates the effective Freidmann equation after inflation. As a consequence, the universe decelerates with $a(t) \sim t^{1/3}$, which is different from the standard radiation model due to the effect of variation in $G$.

The second phase transition takes place at $ \phi = \beta^{-1/2} M_P $ and energy is released  again. It is found that this phase transition (the second one) is a bit different from the first one. At the epoch of the second phase transition, major portion of the energy is released as radiation heating the universe slightly by $ 5.22 \times {10^5}^0 K $ above the temperature $ 246 {\rm GeV} $, which is the temperature of the radiation being red-shifted from the epoch of the first phase transition. An extremely small portion of the released energy causes small fluctuations in the stable state $ \phi = \beta^{-1/2} M_P $. These fluctuations die out soon causing creation of the phantom dark energy. Radiation still dominates dynamics of the universe but with a change during the post-second  phase transition era. In this era, radiation phase expands as $a(t) \sim t^{1/2}$ consistent with the standard radiation model. It is because spontaneous symmetry breaking takes place at the second phase transition and $ G $ becomes a constant equal to $ G_N = M_P^{-2} $. Thus the standard cosmology is recovered at this epoch (the epoch of the second phase transition) with the background radiation having the initial temperature $ \sim 246 {\rm GeV} $.

 This scenario is followed by matter-dominated universe expanding with $ a(t) \sim t^{2/3}.$ The created phantom dark energy dominates the universe at red-shift $ z = 0.46 $. As a consequence, universe speeds up due to a jerk caused by phantom energy.

Natural units ${\hbar} = c = k_B = 1$ (where $\hbar =
  h/2\pi$ with $h$ being the Planck's constant, c being the speed of light and
  $k_B$ being the Boltzmann's constant) are used with GeV as a fundamental
  unit. In these units, it is obtained that $1 {\rm GeV} = 4.96 \times {10^{13}}  ^0 K = 9.07 \times 10^{-24}{\rm gm}$ and $ 1 {\rm GeV}^{-1} = 3.879 \times 10^{-15}{\rm cm} = 1.293 \times 10^{-25}{\rm sec}.$  Here, Newtonian gravitational constant is given as $ G_N = M_P^{-2}$ with $M_P = 2.4\times 10^{18}{\rm GeV}$ being the Planck mass.

\bigskip

\noindent \underline{\bf 2. Early inflation and pre-second phase transition era}

\smallskip

Investigations begin from the action 
$$ S = \int {d^4x} \sqrt{-g} \Big[\frac{\beta\phi^2 R}{16 \pi} - \frac{1}{2}
g^{\mu\nu} \partial_{\mu} \phi \partial_{\nu} \phi - \frac{\lambda (\phi)}{4}|\phi^2 - \beta^{-1}M_P^2|^2\Big], \eqno(1a)$$
where $R$ is the Ricci scalar curvature ,
$$ \lambda(\phi) = K \phi^{-2} \eqno(1b)$$
with constant $K$ having mass dimension 2
and $\partial_{\mu} = {\partial}/{\partial x^{\mu}}.$

Using the condition $\delta S/\delta g^{\mu\nu} = 0,$ action (1a) yields
gravitational equations 
$$ \beta\frac{\phi^2}{16 \pi} [R_{\mu\nu} - \frac{1}{2} g_{\mu\nu} R ] +
\frac{\beta}{16 \pi}[ \phi^2_{;\mu\nu} - g_{\mu\nu} {\Box} \phi^2 ] $$
$$ - \frac{1}{2} \partial_{\mu} \phi \partial_{\nu} \phi - \frac{1}{2}
g_{\mu\nu} \Big[  - \frac{1}{2}
g^{\rho\sigma} \partial_{\rho} \phi \partial_{\sigma} \phi - \frac{\lambda}{4}|\phi^2 - \beta^{-1}M_P^2|^2
 \Big] = 0, \eqno(2a)$$
with
$${\Box} = \frac{1}{\sqrt{-g}} \partial_{\mu} \Big[
\sqrt{-g} g^{\mu\nu} \partial_{\nu} \Big]. \eqno(2b)$$

Taking trace of (2a) and setting $\beta = 8 \pi/3 $ for convenience of calculations, it is obtained that
$$ - \frac{\phi^2}{6} R  - [ \phi {\Box}\phi +
\partial^{\mu} \phi \partial_{\nu} \phi] + \frac{1}{2} \partial^{\mu} \phi
\partial_{\nu} \phi + \frac{\lambda}{2}
\Big|\phi^2 - \frac{3}{8\pi}M_P^2\Big|^2 = 0, \eqno(3)$$

Moreover, using the condition $\delta S/\delta \phi = 0,$ action (1a) yields
 equation for $\phi$ as
$${\Box}\phi +  \frac{\phi}{3} R + \lambda \phi\Big|\phi^2 - \frac{3}{8\pi} M_P^2\Big| + 2 \lambda \phi^{-1}\Big|\phi^2 - \frac{3}{8\pi} M_P^2\Big|^2 = 0.  \eqno(4)$$ 

Eliminating ${\Box}\phi$ from (3) and (4) as well as re-arranging terms, it is
obtained that
$$ \frac{1}{3}R - \phi^{-2} \partial^{\mu} \phi \partial_{\nu} \phi = 2 \lambda \phi^{-2} \Big[- \phi^2 \Big|\phi^2 - \frac{3}{8\pi} M_P^2\Big| - \frac{5}{2}\Big|\phi^2 - \frac{3}{8\pi} M_P^2\Big|^2 \Big] . \eqno(5)$$

Experimental evidences support spatially homogeneous flat model of the
universe \cite{ad}. So, the line-element, giving geometry of the universe, is 
taken as
$$ dS^2 = dt^2 - a^2(t)[dx^2 + dy^2 + dz^2] \eqno(6)$$
with $a(t)$ as the scale factor.

In this space-time, the scalar curvature is obtained as
$$ R = 6 \Big[\frac{\ddot a}{a} + \Big(\frac{\dot a}{a} \Big)^2 \Big]. \eqno(7)$$

Due to spatial homogeneity of the space-time (6), $\phi(x,t) = \phi(t)$ and $ \lambda(\phi) = \lambda(t) $. As mentioned above, in
the proposed model $\phi(t)$ increases with the expansion of the universe. So, it
is reasonable to take solution of (4) as
$$ \phi(t) = \sqrt{\frac{3}{8 \pi}}M_P \Big(\frac{a}{a_*} \Big)^{n}\eqno(8a)$$
with $n$ being a positive real number and $a_*$ being the scale factor when
$\phi = \sqrt{\frac{3}{8 \pi}}|M_P|$ and phase transition takes place, which is discussed below. 

Moreover, $ \lambda(t_0) = \lambda_0$  is proposed to be
$$ \lambda(t) = \lambda_0 \Big(\frac{a_0}{a} \Big)^{2n },   \eqno(8b)$$
where $ a_0 = a(t_0) $ with $ t_0 $ being the present time. 1(b), (8a) and (8b) yield
$$ K = \frac{3}{8 \pi}\lambda_0 M_P^2 \Big(\frac{a_0}{a_*} \Big)^{2n },   \eqno(8c)$$

Connecting (5), (7), (8a) and (8b), we obtain
$$ 2\Big[\frac{\ddot a}{a} + \Big(\frac{\dot a}{a} \Big)^2 \Big] - n^2 
\Big(\frac{\dot a}{a} \Big)^2 = 2 K \Big(\frac{a_*}{a}
\Big)^{4n}\Big[- \Big(\frac{3}{8\pi}\Big)^2 + \frac{3}{2\pi}\Big(\frac{a_*}{a}
\Big)^{2n} - \frac{3}{2}\Big(\frac{a_*}{a}
\Big)^{4n} \Big] . \eqno(9)$$

First integration of (9) yields the Friedmann equation (FE)
$$\Big(\frac{\dot a}{a} \Big)^2 = \frac{C}{a^{(2 + \alpha)}} + 2 K \Big(\frac{a_*}{a}
\Big)^{4n}\Big[- \frac{1}{(\alpha
+ 2 - 4n)} \Big(\frac{3}{8\pi}\Big)^2 $$
$$+ \frac{3}{2\pi(\alpha
+ 2 - 2n)}\Big(\frac{a_*}{a}
\Big)^{2n} - \frac{3}{2(\alpha
+ 2)}\Big(\frac{a_*}{a}
\Big)^{4n} \Big]  , \eqno(10a)$$
where,
$$ \alpha = 2 - n^2 . \eqno(10b)$$
 
In FE (10a), the term ${C}/{a^{(2 + \alpha)}}$
  corresponds to matter density for $\alpha = 1.$ So, to get the matter , which is required for formation of large structure (planned to be discussed in other paper) in a viable cosmology
   $\alpha = 1$ is taken. As a result,
  $$ n  = 1 \eqno(10c) $$
from (10b).

 Further, using $n$ from (10c), (8a), (8b) and (10a) are re-written as
$$ \phi(t) = \sqrt{\frac{3}{8 \pi}}M_P \Big(\frac{a}{a_*} \Big), \eqno(11a)$$
$$ \lambda(t) = \lambda_0 \Big(\frac{a_0}{a} \Big)^{2}\eqno(11b)$$

and
$$\Big(\frac{\dot a}{a} \Big)^2 = \frac{C}{a^3} + 2 K \Big(\frac{a_*}{a}
\Big)^{4}\Big[\Big(\frac{3}{8\pi}\Big)^2 + \frac{3}{2\pi}\Big(\frac{a_*}{a}
\Big)^{2} - \frac{1}{2}\Big(\frac{a_*}{a}
\Big)^{4} \Big] . \eqno(12)$$ 

Thus, it is interesting to see that the matter term $C/a^3$ in (12)emerges spontaneously from the gravitational sector, so it is recognized as dark matter (DM) being non-baryonic. The spontaneous emergence of dark matter is also obtained in references \cite{sks, sks06, sks07, sks08a,  sks08b} from higher-order terms of scalar curvature. The present DM density of the universe is observed as $\rho_{\rm m} = 0.23
\rho^0_{\rm cr}$ \cite{abl} with $$ \rho^0_{\rm cr} = \frac{3 H_0^2}{8 \pi G}= 2.85 \times 10^{-50} {\rm GeV}^4
\eqno(13a)$$
 and 
$$ H_0 = \frac{0.68}{t_0} = 2.035 \times 10^{-43} {\rm GeV}   \eqno(13b)$$
with $t_0 = 13.7 {\rm Gyr} = 6.6 \times 10^{42} {\rm GeV}^{-1}$and $G = M_P^-2, M_P = 2.4\times 10^{18} {\rm GeV}$ in the present time.

Dark matter is characterized as pressureless fluid, so energy conservation equation
$$ \dot \rho  + 3 \frac{\dot{a}}{a}\rho = 0 \eqno(14)$$
yields 
$$ \rho_{\rm m} = \frac{0.69 H_0^2 M_P^2}{8\pi }\Big(\frac{a_0}{a}\Big)^3 \eqno(15a)$$
using observational values given above. The DM density obtained above is
$$ \rho_{\rm m} = \frac{3 \phi^2 C}{8\pi a^3} = \frac{3 M_P^2 C}{8\pi a^3}\Big(\frac{a}{a_*} \Big)^{\sqrt{3/2 \pi}}
 \eqno(15b)$$
using (11a).
At $ a = a_*,$ we obtain
$$  \frac{9 M_P^2 C}{(8\pi)^2 a_*^3} = \frac{0.69 H_0^2 M_P^2}{8\pi }\Big(\frac{a_0}{a_*}\Big)^3
 \eqno(15c)$$
from (15a) and (15b).
(15c) yields the integration constant $C$ as
$$ C = \frac{1.84 \pi  H_0^2 {a_0}^3}{3}. \eqno(15d)$$

Like the DM term, other terms in (12) given as
$$ \rho_G = \frac{3 K}{4 \pi} M_P^2\Big(\frac{a_*}{a}
\Big)^{2}\Big[\Big(\frac{1}{\pi}\sqrt{\frac{3}{2}}\Big)^2 - \Big\{\frac{1}{\sqrt{2}}\Big(\frac{a}{a_*}
\Big)^{2} - \frac{3}{2\pi\sqrt{2}}\Big\}^2 \Big]
 \eqno(16a)$$
also emerge from the gravitational sector. Being caused from the gravitational sector, $ \rho_G $ is recognized as density for the dark energy fluid . 

According to (16), $ \rho_G = 0 $ at 
$ a = a_{**} $ given by
$$ \frac{1}{\sqrt{2}}\Big(\frac{a_{**}}{a_*}
\Big) = \sqrt{\Big(\frac{1}{\pi}\sqrt{\frac{3}{2}}\Big) + \frac{3}{2\pi\sqrt{2}}} \simeq 0.9 .\eqno(16b)$$

Using the pressure density $ p = \omega \rho$ with $ \omega $ being the equation of state (EOS) parameter in (14), 
EOS parameter for the dark energy with density $ \rho_G $ is obtained as
$$\omega_G = - \frac{2}{3} + \frac{4 a}{3 a_*}\frac{\Big\{\frac{1}{\sqrt{2}}\Big(\frac{a}{a_*}
\Big)^{2} - \frac{3}{2\pi\sqrt{2}}\Big\}}{\Big(\frac{a_*}{a}
\Big)^{2}\Big[\Big(\frac{1}{\pi}\sqrt{\frac{3}{2}}\Big)^2 - \Big\{\frac{1}{\sqrt{2}}\Big(\frac{a}{a_*}
\Big)^{2} - \frac{3}{2\pi\sqrt{2}}\Big\}^2 \Big]}. \eqno(16c)$$
(16c) implies $\omega_G < -1/3$ for $ a < a_*.$ It means that this dark energy mimics quintessence.

In the very early universe,
\begin{eqnarray*}
\frac{C}{a^3} &=& \frac{1.84 \pi  H_0^2}{3} \Big(\frac{a_0}{a}\Big)^3 \\ &<& 2  K \Big(\frac{a_*}{a}
\Big)^{4}\Big[\Big(\frac{1}{\pi}\sqrt{\frac{3}{2}}\Big)^2 - \Big\{\frac{1}{\sqrt{2}}\Big(\frac{a}{a_*}
\Big)^{2} - \frac{3}{2\pi\sqrt{2}}\Big\}^2 \Big] .
\end{eqnarray*}
\vspace{-1.8cm}
\begin{flushright}
(17)
\end{flushright}

Using the inequality (17), (12) is obtained as
$$\Big(\frac{\dot a}{a} \Big)^2 = 2  K \Big(\frac{a_*}{a_c}
\Big)^{2} \Big(\frac{a_c}{a}
\Big)^{4} \Big[\Big(\frac{1}{\pi}\sqrt{\frac{3}{2}}\Big)^2 - \Big\{\frac{1}{\sqrt{2}}\Big(\frac{a}{a_c}
\Big)^{2} - \frac{3}{2\pi\sqrt{2}}\Big\}^2 \Big(\frac{a_*}{a_c}\Big)^{2}\Big\}^2\Big] ,\eqno(18)$$
where $ a_c$ is the scale factor at which early inflation ends.
 
(18) integrates to
$$ sech\theta - tanh\theta = D e^{2 \sqrt{K_c}t},           \eqno(19a)$$
where 
$$ \frac{1}{\pi}\sqrt{\frac{3}{2}}                              sech\theta = \frac{1}{\sqrt{2}}\Big(\frac{a}{a_c}
\Big)^{2} - \frac{3}{2\pi\sqrt{2}}\Big\}^2 \Big(\frac{a_*}{a_c}\Big)^{2} \eqno(19b)$$                   
and 
$$ K_c = K \Big(\frac{a_*}{a_c}\Big)^4 = \frac{3}{8 \pi}\lambda_0 M_P^2                    \Big(\frac{a_0}{a_*}\Big)^2 \Big(\frac{a_*}{a_c}\Big)^4.\eqno(19c)$$

(19a) and (19b)yield
$$ \Big(\frac{a}{a_c}
\Big)^{2} - \frac{3}{2\pi}\Big\}^2 \Big(\frac{a_*}{a_c}\Big)^{2} = \Big[ \Big(\frac{a_P}{a_c}
\Big)^{2} - \frac{3}{2\pi\sqrt{2}}\Big\}^2 \Big(\frac{a_*}{a_c}\Big)^{2}\Big]e^{2 \sqrt{K_c}(t - t_P)}                 \eqno(19d)$$
with $ K_c$ given by (19c).

For $ a_P < a \leq a_c << a_*,$ (19d) is approximated to
$$ a \simeq a_P e^{\sqrt{K_c}(t - t_P)},                   \eqno(20)$$
showing the early inflation. 

As $ a << a_* $, the slow-roll parameter $ \epsilon $ is obtained as
$$\epsilon = - \frac{\dot H}{H^2} \simeq \frac{1}{2}, \eqno(21)$$
where $ H = \frac{\dot a}{a}$ is given by (18).

In Refs. \cite{sh, arl}, formula for maximum number of e-fold during inflationary phase was obtained as
$$ N^{\rm max} = 63.3 + 0.25 ln \epsilon \simeq 63.13                             . \eqno(22a$$ 

Connecting (21) and (22a), it is obtained that
$$ \frac{a_c}{a_P}\simeq e^{63.3} = 3.1\times 10^{27}. \eqno(22b)$$ 

\noindent \underline{\bf Exit from inflation and radiation-dominance}

\smallskip

As mentioned above, inflation ends at time $ t = t_{c} ,\phi = \phi_c $ and the first phase transition takes place. At this epoch, an enormous amount of energy is released as latent heat with density
\begin{eqnarray*}
 V(\phi_{P}) - V(\phi_{c}) &= &\frac{1}{4}\lambda_0                     M_{P}^{4}\Big[\Big(\frac{a_0}{a_P}\Big)^2 \Big|\Big(\frac{a_P}{a_*}\Big)^2 - \frac{3}{8\pi}\Big|^2 \\&& - \Big(\frac{a_0}{a_c}\Big)^2 \Big|\Big(\frac{a_c}{a_*}\Big)^2 - \frac{3}{8\pi}\Big|^2 \Big]\\ & \simeq & \frac{1}{4}\lambda_0 M_{P}^{4} \frac{a_0}{a_P}\Big)^2 \Big[ 1 - \Big(\frac{a_P}{a_c}\Big)^2 \Big] .
\end{eqnarray*}
\vspace{-1.8cm}
\begin{flushright}
(23)
\end{flushright}

According to WMAP results \cite{abl}, density of the released energy is obtained  as 
$$V(\phi_P)-(\phi_c) = \frac{3}{8\pi}\times 10^{-18} M_P^4 = 3.96 \times 10^{54}{\rm GeV}^4\eqno(24)$$
at $ \phi = \phi_c $.

The emitted radiation (latent heat) thermalizes the universe after inflation upto the temperature $ T_c = 2.45 \times 10^{14} {\rm GeV}$, which is the GUT (grand unified theory) scale. $ T_c $ is given by 
\begin{eqnarray*}
\frac{1}{4}\lambda_0 M_{P}^{4} \Big(\frac{a_0}{a_P}\Big)^2 \Big[ 1 - \Big(\frac{a_P}{a_c}\Big)^2 \Big] &\simeq & \frac{1}{4}\lambda_0 M_{P}^{4} \Big(\frac{a_0}{a_P}\Big)^2 \\&=& \frac{\pi^2}{15}T_c^4 \\&=& 1.19 \times 10^{57}{\rm GeV}^4. 
\end{eqnarray*}
\vspace{-1.8cm}
\begin{flushright}
(25)
\end{flushright}
using (23)and (24).

Thus energy with radiation density, given by (25), is emitted at the end of inflation and

At the end of inflation, the density of the left - over $ \phi-$ energy is obtained as
\begin{eqnarray*}
 V(\phi_c) = \rho_{rem} &=&  \frac{1}{4}\lambda_0 M_{P}^{4} \frac{a_0}{a_c}\Big)^2 \\ &=& \frac{\pi^2}{15}T_c^4 \Big(\frac{a_P}{a_c}\Big)^2 \\ &=& 0.412 {\rm GeV}^4
 \end{eqnarray*}
\vspace{-2cm}
\begin{flushright}
(26)
\end{flushright}
using (24) and (25).

\noindent This energy is released after the second phase transition at $ a = a_* $ being discussed in the following section. 

The emitted radiation with temperature $ T_c = 2.45 \times 10^{14} {\rm GeV}$ (evaluated below) is given by density
$$ \rho^I_{rd} = \frac{\pi^2}{15}T_c^4 \Big(\frac{a_c}{a}\Big)^4. \eqno(27a)$$ 
Obviously, contribution of this radiation affects the cosmic dynamics leading to modification of the FE (12) as
$$\Big(\frac{\dot a}{a} \Big)^2 = \frac{8\pi}{3 \phi^{2}} \rho_{rd} +  \frac{0.23 H_0^2 {a_0}^3}{a^3} + 2  K \Big(\frac{a_*}{a_c}
\Big)^{2} \Big(\frac{a_c}{a}
\Big)^{4} \Big[\Big(\frac{1}{\pi}\sqrt{\frac{3}{2}}\Big)^2$$
$$ - \Big\{\frac{1}{\sqrt{2}}\Big(\frac{a}{a_c}
\Big)^{2} - \frac{3}{2\pi\sqrt{2}}\Big\}^2 \Big(\frac{a_*}{a_c}\Big)^{2}\Big\}^2\Big] \eqno(27b)$$

Using (11a), (11b), (25) and (27a), it is obtained that at $  a = a_c,$
$$ \frac{8\pi}{3 \phi^{2}} \rho^I_{rd} > 2  K \Big(\frac{a_*}{a_c}
\Big)^{2} \Big(\frac{a_c}{a}
\Big)^{4} \Big[\Big(\frac{1}{\pi}\sqrt{\frac{3}{2}}\Big)^2 - \Big\{\frac{1}{\sqrt{2}}\Big(\frac{a}{a_c}
\Big)^{2} - \frac{3}{2\pi\sqrt{2}}\Big\}^2 \Big(\frac{a_*}{a_c}\Big)^{2}\Big\}^2\Big] .\eqno(27c)$$

Using inequalities (17) and (27c), (27b) reduces to
$$\Big(\frac{\dot a}{a} \Big)^2 \simeq \frac{8\pi^{3}}{45 M_P^{2}}T_c^4 \frac{a_c^4 a_*^2}{a^6}\eqno(27d)$$
incorporating (11b) and (27a). This equation is integrated to
$$ a(t) = a_c \Big[1 + 3 \sqrt{\frac{8\pi^{3}}{45 }}\frac{T_c^2 a_c^2 a_*}{M_P}(t - t_c) \Big]^{1/3} \eqno(29)$$

 The result (29) justifies exit of the early universe from inflationary phase at $t = t_c.$ It also shows deceleration for $t > t_c.$ 

Here, it is important to remark that the expansion (29) is different from the standard radiation model due to the effect of variable $ G = \frac{3}{8 \pi} \phi^{-2}.$

 \bigskip

\noindent \underline{\bf 3. Post-second phase transition era}

\smallskip

It is discussed above that the first phase transition takes place at $ \phi = \phi_c $. But this state is not stable. So, $ \phi $ rolls down the potential hill $ V(\phi) $ to a stable state being obtained from the condition  $dV(\phi)/d\phi = 0 $ . For
the potential $V(\phi)$ in action (1), this condition looks like
$$ (\phi^2 + \frac{3}{8\pi} M_P^2) \Big|\phi^2 - \frac{3}{8\pi} M_P^2 \Big| = 0                                          \eqno(30)$$ 

(30) yields stable fixed points
$$ \phi = \pm \sqrt{\frac{3}{8\pi}}M_P .\eqno(31)$$

Spontaneous symmetry breaking takes place, when $ \phi $ reaches the stable states given by (31). As a consequence, $  G$ acquires a constant value $ M_P^{-2} = G_N $ (the Newtonian gravitational constant). Moreover, the second phase transition takes place and energy with density $ V(\phi_c) $ is released at $ \phi = \sqrt{\frac{3}{8\pi}}|M_P |$ . In what follows, it is discussed that the major portion of this energy is released as radiation, but an extremely small portion causes  fluctuations in the state $ \phi = \sqrt{\frac{3}{8\pi}}|M_P| $ dying out in a very short interval of time.

So, $\phi$ is taken as 
$$\phi = \sqrt{\frac{3}{8\pi}}M_P + \delta\phi, \eqno(32)$$ 
where $\delta\phi$ is a small fluctuation in the state $\phi = \sqrt{\frac{3}{8\pi}}|M_P|.$

Using (32) in (3) and (4), it is obtained that
$$ \frac{(M_P^2 + \sqrt{\frac{3}{2 \pi}} M_P \delta\phi)}{6 } R  + \frac{1}{2}\sqrt{\frac{3}{ \pi}} M_P {\Box}\delta\phi  = 0 \eqno(33)$$
and
$${\Box}\delta\phi +  \frac{1}{3} R \delta\phi + 2K \delta\phi = 0  \eqno(34)$$ 
neglecting higher orders of $ \delta\phi.$

In the space-time (6),(34) is re-written as
$${\ddot \delta\phi} +  3 \frac{\dot a}{a}{\dot \delta\phi} + \Big[ 2\Big\{\frac{d}{dt} \Big(\frac{\dot a}{a}\Big) + 2 \Big(\frac{\dot a}{a}\Big)^2 \Big\} + 2 K \Big] \delta\phi  = 0.  \eqno(35)$$
 
(35) is an equation of damped harmonic oscillator
 having decreasing amplitude  with increasing scale factor $ a(t)$, so a trial solution can be taken as
$$ \delta\phi = A \Big[\Big(\frac{ a_*}{a}\Big)^r - \Big(\frac{ a_*}{a_e}\Big)^r \Big] cos [\varpi(t - t_*)]  \eqno(36)$$
where $r$ is a positive real number, $ \varpi $ is the frequency and $ t_* $ is the phase transition time. Here $a_e << a_0$ is the scale factor, when fluctuation vanishes.

(36) satisfies (35) if 
$$ r = \frac{3}{2},  \quad \varpi = \sqrt{\frac{K}{\alpha}} \eqno(37a,b)$$
with $ \alpha $ being a constant and
$$ \Big(\frac{\ddot a}{a}\Big) +  \frac{5}{2} \Big(\frac{\dot a}{a}\Big)^2 = 2(\varpi^2 - 2 K).  \eqno(37c)$$

Connecting (7),(33) and (36), we obtain 
$$\frac{\ddot a}{a} +  \Big(\frac{\dot a}{a}\Big)^2 = \frac{2\varpi^2}{M_P}\sqrt{\frac{2 \pi}{3}} \delta\phi.     \eqno(38)$$ 
after some manipulations.

(37c) and (38) yield
\begin{eqnarray*}
\Big(\frac{\dot a}{a}\Big)^2 = \Big|\Big(\frac{\dot a}{a}\Big)^2 \Big|&=& \frac{4}{3}(\varpi^2 - 2 K) - \frac{4\varpi^2}{3 M_P}\sqrt{\frac{2 \pi}{3}} |\delta\phi |\\&=& \frac{4}{3} \Big[(\varpi^2 - 2 K) - \frac{\varpi^2}{ M_P}\sqrt{\frac{2 \pi}{3}} A \Big\{\Big(\frac{ a_*}{a}\Big)^{3/2}- \Big(\frac{ a_*}{a_e}\Big)^{3/2}\Big\}\Big] \\&=& \frac{4K}{3\alpha} \Big[(1 - 2 \alpha) -  \frac{A}{M_P}\sqrt{\frac{2\pi}{3}} \Big\{\Big(\frac{ a_*}{a}\Big)^{3/2}- \Big(\frac{ a_*}{a_e}\Big)^{3/2}\Big\}\Big]
\end{eqnarray*}
\vspace{- 1.8cm}
\begin{flushright}
(39)
\end{flushright}
using (8c), (10c), (36) and (37a,b). 

The kinetic energy density of the oscillating $ \delta\phi $ is obtained as
\begin{eqnarray*}
\rho_{\delta\phi} &=& \frac{1}{2}\Big[ - \frac{3}{2}\Big( \frac{\dot a}{a} \Big) + \varpi tan [\varpi(t - t_*)] \Big]^{2} {\delta\phi }^{2}\\& \simeq& \frac{9}{8}\Big( \frac{\dot a}{a} \Big)^{2}A^2 \Big\{\Big(\frac{ a_*}{a}\Big)^{3/2}- \Big(\frac{ a_*}{a_e}\Big)^{3/2}\Big\}^2.
\end{eqnarray*}
\vspace{- 1.8cm}
\begin{flushright}
(40)
\end{flushright}
using (36) and $ |cos [\varpi(t - t_*)]| \leq 1.$

(40) shows that kinetic energy of $ \delta\phi $ decreases with growing $ a(t).$ It means that during oscillation, $ \delta\phi$ emits energy, which is being pumped to the universe. The energy density in the universe caused by fluctuations around the stable state $ \phi = \sqrt{\frac{3}{8\pi}}|M_P| $
is given by the energy density 
$$ \rho_{phde}= \frac{K M_P^2}{2\alpha \pi} \Big[(1 - 2 \alpha) -  \frac{A}{M_P}\sqrt{\frac{2\pi}{3}}\Big\{\Big(\frac{ a_*}{a}\Big)^{3/2}- \Big(\frac{ a_*}{a_e}\Big)^{3/2}\Big\} \Big] \eqno(41)$$

As phantom dark energy is seeded by decay of $ \delta\phi$, so $\rho_{phde} = 0$ at $ a = a_* , \rho_{phde}= 0$ at $a = a_*$. Using this condition in (41)
$$ (1 - 2 \alpha) =  \frac{A}{M_P}\sqrt{\frac{2\pi}{3}}\Big\{1 - \Big(\frac{ a_*}{a_e}\Big)^{3/2}\Big\}                        \eqno(42)$$

Now, (8c), (10c),(41) and (42) yield
$$\rho_{phde}= \frac{\lambda_0 A M^3_P}{2\alpha \pi}                                \sqrt{\frac{2\pi}{3}}\Big(\frac{a_0}{a_*}\Big)^2\Big\{1 - \Big(\frac{ a_*}{a}\Big)^{3/2}\Big\}                       \eqno(43)$$

When $ \rho_{phde}$, given by (43), satisfies the conservation equation (14), the equation of state parameter $w_{phde}$ is obtained as
$$ w_{phde} = - 1 - \dfrac{\Big(\frac{ a_*}{a}\Big)^{3/2}}{2 \Big[1 - \Big(\frac{ a_*}{a}\Big)^{3/2}\Big]}                            . \eqno(44a)$$

The result (44a) implies that the energy emitted due to decay of $ \delta\phi$ mimics phantom dark energy as $ w_{phde} < - 1. $

Further, according to observations, dark energy density of the present universe  $\rho^0_{phde} = 0.73 \rho^0_{\rm cr} $ is extremely low and phantom energy density increases with the expansion of the universe. So, phantom energy density is expected much lower than 
$\rho^0_{phde}$ for $ a < a_0 $. It shows that an extremely small portion  of released energy with density 
$$\varrho = 10^{-x} V(\phi_c) = 0.412 \times 10^{-x} {\rm GeV}^4    \eqno(44b)$$ 
is converted into phantom dark energy and the major portion $ \approx 0.412{\rm GeV}^4 $ is emitted as radiation at the second phase transition like latent heat. Here $x$ is a positive real number.

As discussed above, the emitted radiation after the first phase transition has temperature $246 {\rm GeV}$ after red-shifting to the scale factor $ a_* > a_c $. Using these value in the formula for temperature in the expanding universe, we obtain
$$ \frac{a_*}{a_c} = \frac{T_c}{246} = 10^{12}.       \eqno(45a)$$

Connecting (8a) and (45a), it is obtained that
$$\phi_c = \sqrt{\frac{3}{8\pi}}\times 10^{-12}M_P = 3.45    \times 10^{-13}M_P. \eqno(45b)$$

 Moreover, at the second phase transition, radiation is emitted having density given by (26). Now, at this epoch, radiation is available from two sources (i) red-shifted radiation emitted at first phase transition having temperature $ 246{\rm GeV} $ and (ii) radiation being emitted at the second phase transition. So, total radiation density at $a = a_*$ is obtained as
 \begin{eqnarray*}
  \rho^{*}_{\rm rd} &=&\frac{\pi^2}{15}T_*^4 \\&=& \frac{\pi^2}{15}\times (246)^4 + 0.412 .   
 \end{eqnarray*}
\vspace{- 1.8cm}
\begin{flushright}
(46a)
\end{flushright} 
  
  (46a) yields
 $$ T_* = 246 [1 + 4.275\times 10^{-11}]{\rm GeV} = 246{\rm GeV} + 5.22 \times {10^{5}}^{0} K. \eqno(46b)$$ 
  
 This result shows that ,after the second phase transition, the universe is reheated slightly by temperature $5.22\times {10^5 }^{0} K$. 
 
 So, in this era, radiation density is obtained as
 $$ \rho^{II}_{\rm rd} = \frac{\pi^2}{15}T_*^4 \Big(\frac{a_*}{a} \Big)^4. \eqno(46c)$$
  
 As the released energy causes fluctuation in the state $  \phi = \sqrt{\frac{3}{8\pi}} |M_P|$, $ \delta\phi$ has energy with density
$$ \rho_{\delta\phi}  = 0.412\times 10^{-x} {\rm GeV}^4 \eqno(47)$$
 at $ a = a_* $. This result is obtained using (44b).

\noindent \underline{\bf Radiation-dominance during post-second phase transition era}

 In this era too, radiation-dominated universe is obtained ,but with a
changed scenario. In this case, Friedmann equation (27d) gets modified again and the variable $G$ is replaced by the constant $G = G_N = M_P^{-2}$ leading to the
standard form of FE during radiation-dominance. 
 The modified (27d) is obtained as
 \begin{eqnarray*}
\Big(\frac{\dot a}{a} \Big)^2 &= &\frac{8\pi^3}{45 M_P^{2}}T_*^4 \Big(\frac{a_*}{a} \Big)^4
\\ &=& 3.56\times 10^{-27}\Big(\frac{a_*}{a} \Big)^4 
\end{eqnarray*}
\vspace{- 1.8cm}
\begin{flushright}
(48)
\end{flushright}
 using (25),(45a) and (46c).
 
 (48) is integrated to 
 $$a = a_* [ 1 + 1.19 \times 10^{-13}(t - t_*)]^{1/2}                                \eqno(49)$$
 
 It shows that the standard model of cosmology is recovered after the second phase transition, so emitted radiation with temperature $ T_* = 246 {\rm GeV} + 5.22 \times {10^{5}}^{0} K$ is identified as the cosmic microwave background (CMB)radiation having temperature $ T_0 = {2.73}^0K$ in the present universe.
 
 Using $T_0$ and (46b), it is obtained that
 $$ \frac{a_0}{a_*} = \frac{T_*}{T_0} = 4.47 \times 10^{15}                        \eqno(50a)$$                    
                                                   
(45a)and (50a) yield
$$ \frac{a_0}{a_c} = \frac{a_0}{a_*} \frac{a_*}{a_c}= 4.47 \times 10^{27}                       \eqno(50b)$$ 
as well as
(26) and (50b) imply
$$\lambda_0 = 7.44 \times 10^{-127}.                           \eqno(50c)$$

 After knowing $ \lambda_0 $ from (50c) as well as connecting (19c), (20), (22b), (45a) and (50a), inflationary period (in the pre-second phase transition era) is evaluated as
 $$ (t_c - t_P) = 4.41 \times 10^{7}{\rm GeV}^{-1} = 5.7 \times 10^{-18} {\rm sec}.  \eqno(51)$$

It is discussed above that extremely portion of the energy having density $ \varrho $ of $ \delta\phi $ decays to phantom dark energy, so
$$ - {\dot \varrho} =                                      {\dot \rho_{phde}} . \eqno(52a)$$ 

As phantom is seeded by decay of $ \delta\phi$, so  at $ a = a_* ,\rho_{phde} = 0 $ . Using this condition as well as (47), (52a) is integrated to
$$ 0.412\times 10^{-x}  - \rho_{\delta                                \phi}  =  \rho_{phde} .\eqno(52b)$$     

Using (40), (42), (48) and the condition  at $ a = a_* , \alpha $ is obtained that
$$ (1 - 2\alpha)^2 = \frac{2}{9}\times \frac{0.412\times 10^{-x}}{T^4_*}  .                                    .\eqno(53)$$   

According to (40), $ \rho_{\delta\phi} = 0 $ at $a = a_e$, so we obtain
$$\frac{\lambda_0 A M^3_P}{2\alpha \pi}                                \sqrt{\frac{2\pi}{3}}\Big(\frac{a_0}{a_*}\Big)^2\Big\{1 - \Big(\frac{ a_*}{a_e}\Big)^{3/2}\Big\}                      = 0.412 \times 10^{-x} \eqno(54)$$
using (43). 

Connecting (42),(53) and (54), a quadratic equation for $(1 - 2\alpha)$ is obtained as  
$$ (1 - 2\alpha)^2 - (1 - 2\alpha) + \frac{2\lambda_0  M^4_P}{\pi T^4_*}\Big(\frac{a_0}{a_*}\Big)^2 = 0. \eqno(55a)$$
(55a) yields
$$ (1 - 2\alpha) \simeq \frac{2\lambda_0  M^4_P}{\pi T^4_*}\Big(\frac{a_0}{a_*}\Big)^2  = 8.53 \times 10^{-32}       \eqno(55b)$$
using (50a) and (50c).

Further (53) and (55b) imply
$$ 10^{-x}= 2.95 \times 10^{-52}       \eqno(55c)$$

The present dark energy density is
$$ \rho^0_{phde} = 0.73 \rho^0_{\rm cr} = 0.73\times \frac{3 H_0^2 M_P^2}{8 \pi} = 2.08 \times 10^{-50}{\rm GeV}^4 .      \eqno(56)$$

So, from (43) and (55), it is obtained that
 $$\frac{\lambda_0 A M^3_P}{2\alpha \pi}                                \sqrt{\frac{2\pi}{3}}\Big(\frac{a_0}{a_*}\Big)^2\Big\{1 - \Big(\frac{ a_*}{a_0}\Big)^{3/2}\Big\}                      = 2.08 \times 10^{-50}{\rm GeV}^4.\eqno(57)$$

(43), (53).(54),(55c) and (57) imply
$$\frac{\rho^0_{phde}}{\rho^e_{phde}}=\dfrac{\Big\{1 - \Big(\frac{ a_*}{a_0}\Big)^{3/2}\Big\}}{\Big\{1 - \Big(\frac{ a_*}{a_e}\Big)^{3/2}\Big\}}$$
$$=\dfrac{2.08 \times 10^{-50}}{0.412\times 10^{-x}}= 171.137 .\eqno(58)$$

Using (50a) for ${ a_*}/{a_0}$ in (58), it is obtained that
$$  \Big(\frac{ a_*}{a_e}\Big)^{3/2} = 1 - \frac{1}{171.137} \Big[1 - \Big(\frac{ a_*}{a_0}\Big)^{3/2}\Big]                           $$
$$ \approx .994 . \eqno(59a)$$
(59) shows that 
$$ a_e = 1.004 a_*. \eqno(59b)$$

Thus
$$\rho_{phde} = 2.08\times 10^{-50}\dfrac{\Big\{1 - \Big(\frac{ a_*}{a}\Big)^{3/2}\Big\}}{\Big\{1 - \Big(\frac{ a_*}{a_0}\Big)^{3/2}\Big\}} $$
$$ = 2.026\times 10^{-50} \Big\{1 - \Big(\frac{ a_*}{a}\Big)^{3/2}\Big\} .\eqno(59c)$$
using (56),(58) and (59b).

This result yields
$$\rho_{phde} = \rho^0_{phde}\Big[1 - 0.9747\Big\{1  - \Big(\frac{ a_*}{a}\Big)^{3/2}\Big\} \Big]\eqno(59d)$$
for $ a < a_0 $ and 
$$\rho_{phde} = \rho^0_{phde}\Big[1 + 0.9747\Big\{1  - \Big(\frac{ a_*}{a}\Big)^{3/2}\Big\} \Big]\eqno(59e)$$
for $ a > a_0 $ using (50a), (56), (59a) and (59c).

(44a) and (50a) imply
$$ w^0_{phde} = - 1 - \dfrac{\Big(\frac{ a_*}{a_0}\Big)^{3/2}}{2 \Big[1 - \Big(\frac{ a_*}{a_0}\Big)^{3/2}\Big]} $$
$$ = -1 - 3.335\times 10^{-24}.                          \eqno(60)$$
This equation shows that EOS parameter $w^0_{phde}$ for phantom energy very much closed to $-1$ in the present universe. 
 
 Using (44a), (59d) and (59e), $ w_{phde}$ and                   $\rho_{phde}$ are tabulated below.
 
 \smallskip
 \centerline{\textbf{Table no.1}}

\centerline{$w$ and $ \rho_{phde}$ are tabulated for increasing $a(t)$ }
\vspace{0.5cm}
 
\begin{tabular}{||c||c||c||}
\hline $a_0/a$ & $w$ & $\rho_{phde}$    \\
\hline $ 4.45 \times 10^{15} $ & $ -83.83 $ & $ \rho^0_{phde}/171.137 $ \\ 
\hline $ 2.8 \times 10^{15} $ & $ -2$ & $ \rho^0_{phde}/2.05$  \\ 
\hline $ 10^{15} $ & $ -1.06$  & $ \rho^0_{phde}/1.15$  \\ 
\hline $ 1091.98 $ & $ -1 - 0.0605\times 10^{-18} $ & $\rho^0_{phde}[1.9747 - 0.118 \times 10^{-18}]  $ \\ 
\hline $ 1.46 $ & $ -1 - 2.95\times 10^{-24} $ & $\rho^0_{phde}[1.9747 - 5.9 \times 10^{-24}]  $ \\ 
\hline $ 1 $ & $ -1 - 1.673\times 10^{-24} $ & $\rho^0_{phde}$ \\ 
\hline $ 0 $ & $ -1 $ & $1.9747 \rho^0_{phde} $ \\ 
 \hline  
\end{tabular} 

\smallskip
 Table no.1 shows that $ w_{phde} = -1 $ for                       
$ a \to \infty. $ This is the EOS parameter for the vacuum. It means that phantom dark energy will evaporate and contribute to vacuum energy with density $1.9747 \rho^0_{phde} $ as $ a \to \infty. $ Moreover, it is interesting to see that $\rho_{phde}$ grows very fast during radiation-dominance compared to the era of dark matter-dominance and phantom-dominance.

\noindent \textbf{\underline{Matter-dominance}} It is obtained above that, even after the second phase transition, radiation dominates the universe. As dark matter density $ \rho_{m} $ given by (15b)and (15d)as well as radiation density (46b) decreases with growing $ a(t)$, a stage is reached at $ a = a_d > a_*, $ such that
$$ \frac{\pi^2}{15}T^4_* \Big(\frac{a_*}{a_d} \Big)^4 = \frac{\pi^2}{15}T^4_* \Big(\frac{a_*}{a_0} \Big)^4\Big(\frac{a_0}{a_d} \Big)^4 $$
$$= 0.23 \rho^0_{\rm cr} \Big(\frac{a_0}{a_a} \Big)^3 = 0.66 \times 10^{-50}\Big(\frac{a_0}{a_d} \Big)^3 \eqno(61a)$$
using (13a). 

Thus, $ \rho^{II}_{\rm rd} < \rho_{\rm m}$ for $ a > a_d$.

Using (50a) and $ T_* = 246{\rm GeV} + 5.22 \times 10^{5} {0}^K$ in (61a), it is obtained that
$$ \frac{a_0}{a_d} = 1092.98  .    \eqno(61b)$$
 It shows the beginning of induced dark matter-dominance at red-shift
 $$  z_d = \frac{a_0}{a_d} - 1 = 1091.98. \eqno(61c)$$                                  
 
This value of red-shift is very much closed to the red-shift $ z = 1087^{+1}_{-2} $ for decoupling of matter from radiation according to WMAP observations.

So, for $ a > a_d,$ (27b) reduces to
$$\Big(\frac{\dot a}{a} \Big)^2 =  \frac{0.23 H_0^2 {a_0}^3}{a^3} \eqno(62a)$$
using (16a) and (16b).

(62a) yields the solution
$$ a = a_d [ 1 + 0.587 H_0 (1 + z_{d})^{3/2} (t - t_{d})]^{2/3} \eqno(62b)$$ 
showing deceleration.

It is interesting to see that, universe expands like standard matter-dominated model.

\noindent \textbf{\underline{Late acceleration}} It is discussed above that density of the phantom dark energy caused by fluctuation around the stable state $ \phi = \sqrt{\frac{3}{8\pi}} |M_P|$ grows and matter density decreases with the expansion of the universe. So, it is obvious for $ \rho_{phde}$ to overpower $ \rho_{\rm m} $ on sufficient growth of $ a(t)$. If this transition takes place at $ a = a_{tr} $, 
$$ 0.6555\times 10^{-50}\Big(\frac{ a_0}{a_{tr}}\Big)^3 = 2.0273 \times 10^{-50}\Big\{1 - 3.33\times 10^{-24}\Big(\frac{ a_0}{a_{tr}}\Big)^{3/2}\Big\}. \eqno(63)$$ 
using (13a),(16a), (16b) and (59c).
(63) yields
$$ \frac{ a_0}{a_{tr}} \simeq 1.46 . \eqno(64)$$ 

It means that phantom dark energy dominates for red-shift $ z < 0.46 $ and the effective Friedmann equation is obtained as
$$\Big(\frac{\dot a}{a} \Big)^2 = \frac{8\pi}{3M_P^2}\rho_{phde} $$
$$= 3.015 \times 10^{-86}\Big\{1 - \Big(\frac{ a_*}{a}\Big)^{3/2}\Big\} \eqno(65a)$$     
using (39),(42),(58) and (59c).

On solving (65a), it is obtained that
$$a(t)= a_* cosh^{4/3}[27.09 + 2.315\times 10^{-43}(t - t_{tr})]$$
$$ = 3.26 \times 10^{-15}a_{tr}cosh^{4/3}[27.09 + 2.315\times 10^{-43}(t - t_{tr})]\eqno(65b)$$ 
using (50a) and (64).

(65b) shows acceleration for $t > t_{tr}.$ It is interesting to see $ a(t) $, given by (65b), will reduce to de-Sitter like expansion asymptotically.

Further using (50a) and $ a_0 = a(t_0) $ in (65b), it is obtained that
$$ t_{tr} \simeq 0.183 t_0.  \eqno(65c)$$

\smallskip
\centerline \textbf{\underline{4. Summary}} Here, cosmology is obtained from the variable gravitational constant and its fluctuations around the stable state $ G = G_N .$ Variation in $ G $ is manifested by a scalar $ \phi(x) $ given as $ \phi = \sqrt{3/8\pi} G^{-1/2} $. The theory is based on $ \phi $ and the Ricci scalar curvature $ R $ only. So, the action contains Einstein-Hilbert like term with $ G $ depending on $ \phi $ as well as kinetic energy and self-interaction potential term for it. The model of the universe begins at the Planck scale being the fundamental scale.

This model provides an interesting cosmology. It is found that trace of the gravitational field equations yields the Friedmann equation with dark matter and a fluid term behaving as quintessence dark energy. These terms emerge as combined effect of $ \phi $ and $ R $. Here, it is investigated that the very early universe inflates for an extremely small period of time being $ 5.7\times 10^{-18}{\rm sec} $ and exits from the inflationary phase at $ t_c = t_P + 5.7\times 10^{-18}{\rm sec} (t_p = M_P^{-1} $ is the Planck time), when $ \phi = \phi_c = \sqrt{3/8\pi}\times 10^{-12}M_P $. A phase transition takes place at this epoch and a huge amount of energy is released as latent heat thermalizing the universe upto the temperature $ 2.45\times 10^{14}{\rm GeV} $ After the inflationary phase, universe decelerates driven by emitted radiation. But, due to the effect of variable $ G $, $ a(t) \sim t^{1/3} $, which is not consistent with standard radiation-model.

After sufficient expansion, when temperature of the emitted radiation red-shifts to the temperature $ 246 {\rm GeV} $ (the electro-weak scale), another phase transition takes place when $ \phi $ rolls down to the stable state $ \phi = \sqrt{3/8\pi}|M_P|.$ Major portion of the released energy reheats the universe slightly and the temperature is raised by $ 5.22 \times 10^5 {0}^K $ and an extremely small portion of the released energy of the order of $ 2.95\times 10^{-52}{\rm GeV}^4 $ is converted to dark energy caused by fluctuations around the stable state $ \phi = \sqrt{3/8\pi}|M_P|.$ Thus, after the second phase transition, $ G $ is no longer a variable and acquires the constant value $ G = G_N $. Radiation still dominates the cosmic dynamics. But, in the post-second phase transition era, radiation-dominated phase decelerates $ \sim t^{1/2} $ being consistent with the standard model. Thus, the standard model of cosmology is recovered at the epoch of the second phase transition and the radiation emitted at this epoch is identified as the cosmic background radiation having the present temperature $ 2.73^0 K. $ Thus, in this model, standard cosmology begins with temperature $ 246 {\rm GeV} + 5.22 \times {10^5 }^{0}K.$ It happens so at the red-shift $ z = 4.47\times 10^{15}.$ 

As mentioned above, $ \phi $ fluctuates around the stable state $ \phi = \sqrt{3/8\pi}$  $\times |M_P|$ after the second phase transition, which oscillates for a very small period till $ a \simeq 1.004 a_* $. It is found that phantom dark energy is seeded by the released energy during fluctuation, which grows later (after fluctuation ends) in the expanding universe such that the present dark energy density is 171.137 times the phantom dark energy at $ a \simeq 1.004 a_* $. This phenomenon takes place during radiation-dominance.

As radiation density falls faster than the dark matter density, dark matter dominates the radiation density at red-shift $ \sim 1091.98 $. It is interesting to see that this red-shift is very closed to the red-shift $ 1098^{+1}_{-2} $ for decoupling of matter from radiation as per WMAP observations \cite{abl}. During the dark matter-dominance, the universe decelerates as $ \sim t^{2/3} $, which is consistent with standard matter-dominated model.

It is discussed above that the phantom dark energy grows during radiation-dominance and matter-dominance. The dark matter density decreases in the expanding universe. So, obviously on sufficient growth of $ a(t) $, dominance of phantom over dark matter is expected. In the present scenario, it is found that phantom overpowers dark matter at red-shift $ z_{tr} = 0.46,$ which is within the range $ 0.46 \pm 0.13$ obtained by 16 Type Supernovae observations \cite{ag}. This transition takes place at time $ t_{tr} = 0.13 t_0.$ At this epoch, phantom energy with growing density in expanding universe gives a jerk. As a consequence, the decelerating universe is accelerated.   

Interestingly, EOS parameter $ w_{phde} $ for phantom energy, created by fluctuation of $ \phi $, depends on the scale factor $ a(t) $. It is obtained that $ w_{phde} \to -1$ and $ \rho_{phde} \to 1.9747\rho^{0}_{phde} $as $ a \to \infty.$ Also, $ a(t) $ reduces to the de-Sitter form asymptotically. These results show that ultimately phantom energy will evaporate contributing to the vacuum energy.

 The following are salient features of this model.
 
 \noindent (i)Quintessence dark energy in the early universe, visible radiation, dark matter and phantom dark energy are caused by the scalar field $ \phi = \sqrt{3/8\pi}G^{-1/2} $. Moreover, phantom dark energy is created due to fluctuations of $ \phi $ around the stable state $ \phi = \sqrt{3/8\pi}|M_P| $.
 
\noindent (ii) Two phase transitions take place in this cosmological picture. The first one takes place at the GUT scale when the early universe comes out of the inflationary phase and the second one takes place at electro-weak scale. This epoch heralds the standard model of the universe with background radiation having temperature $ \sim 246 {\rm GeV}$ with its remnant having temperature $ 2.73^{0}K $ in the present universe.

\noindent (iii) Here, it is shown that phantom dark energy is created by fluctuations in the stable state $ \phi = \sqrt{3/8\pi}M_P = \sqrt{3/8\pi}G_N^{-1/2}$. Thus, this model provides a possible way of creation of phantom energy almost in the beginning of the mtter-dominated phase of the universe. It is found that growth of phantom density is very fast during radiation-dominance in the post-second phase transition era, but it slows down later on. It is found that phantom density will become almost double of the present dark energy density, when $ a \to \infty $. It is interesting to obtain that phantom density will evaporate and will contribute to vacuum energy ultimately.

 \noindent (iv) The scale factor $ a(t) $ is free from big-rip problem. 

Thus, this model yields a physically viable and unified cosmological picture from early universe to the future universe.Results are consistent with observations. Most interestingly, a possible way for creation of phantom dark energy is suggested here.

\end{document}